\documentclass{aa}  
\usepackage{color} 

\usepackage{graphicx}
\usepackage{subfigure}
\usepackage{txfonts}
\usepackage[pdftex]{hyperref}
\begin{document}

   \title{On the formation of hydrogen-deficient low-mass white dwarfs}


   \author{Tiara Battich\inst{1,2}, Leandro G. Althaus\inst{1,2} \& Alejandro H. C\'orsico\inst{1,2} }

   \institute{Instituto de Astrof\'isica de La Plata, CONICET--UNLP, Argentina
  		\and     
     		Facultad de Ciencias Astron\'omicas y Geof\'{\i}sicas, UNLP, Argentina \\
            \email{tbattich@fcaglp.unlp.edu.ar}}

   \date{Received M D, Y; accepted M D, Y}
\authorrunning{Battich et al.}
 
  \abstract
   {Two of the possibilities for the formation of low-mass ($M_{\star}\lesssim 0.5\,M_{\odot}$) hydrogen-deficient white dwarfs are the occurrence of a very-late thermal pulse after the asymptotic giant-branch phase or a late helium-flash onset in an almost 
   stripped core of a red giant star.}
   {We aim to asses the potential of asteroseismology to distinguish between the hot flasher and the very-late thermal pulse scenarios for the formation of low-mass hydrogen-deficient white dwarfs.}
   {We compute the evolution of low-mass hydrogen-deficient white dwarfs from the zero-age main sequence in the context of the two 
   evolutionary scenarios. We explore the pulsation properties of the resulting models for effective temperatures characterizing the instability strip of pulsating helium-rich white dwarfs.}
   {We find that there are significant differences in the periods and in the period spacings associated with low radial-order ($k\lesssim 10$) gravity modes for white-dwarf models evolving within the instability strip of the hydrogen-deficient white dwarfs.}
   {The measurement of the period spacings for pulsation modes with periods shorter than $\sim500\,$s may be used to distinguish between the two scenarios. Moreover, period-to-period asteroseismic fits of low-mass pulsating hydrogen-deficient white dwarfs can help to determine their evolutionary history.}


   \maketitle
%

\section{Introduction}
White dwarf (WD) stars are the final stage in the life of the vast bulk of stars. 
Among the WDs, a great majority ($\sim 80$\%) presents hydrogen (H)-rich atmospheres (DA WDs). However, there is a significant amount of stars ($\sim 20$\%) with H-deficient surfaces. H-deficient WDs exhibit a variety of spectral classes. Among them, there are the helium (He)-rich DO WDs (with effective temperatures, $T_{\rm eff}$, in the range $45000\,{\rm K}\lesssim T_{\rm eff}\lesssim 20\,0000\,{\rm K}$), the He-rich DB WDs ($11000\,{\rm K}\lesssim T_{\rm eff}\lesssim 45000\,{\rm K}$, with only few stars found in the range $30\,000\,{\rm K}\lesssim T_{\rm eff}\lesssim 45000\,{\rm K}$), and WDs with mainly carbon (C) or oxygen (O) in their spectra. The formation channel of these H-deficient WDs was for decades, and still is, a matter of study (see, for instance, \citealt{1979ASSL...75..155R,1979A&A....79..108S,1981ASSL...89..319R,1983ApJ...264..605I} for earlier discussions on this matter). As the population of H-deficient WDs presents a variety of spectral types -- among other particularities -- , it may be fed by different formation channels \citep{2010A&ARv..18..471A}. In particular, DO and DB WDs are believed to be mostly the progeny of PG1159 stars, 
which are hot C-, O-, and He-rich WDs and pre-WDs. These stars, in turn, are expected to form via the very-late thermal-pulse scenario (VLTP) \citep{2006A&A...449..313M}. 

In the VLTP scenario, a post-AGB star experiences a final thermal pulse when the H-burning shell is almost extinct. Therefore, due to a low entropy barrier in this almost extinct H-burning shell -- see \citealt{1976ApJ...208..165I} -- , convective processes carry H to the hot He-burning shell. As a consequence, all -- or almost all -- the H is burned. As diffusion processes take place, a PG1159 star would evolve to a DO WD first, and to a DB WD later \citep{2005A&A...435..631A}. DO WDs could also be the descendants of low-mass H-deficient supergiants R Corona Borealis stars (RCrBs) -- possibly linked to the O(He) stars -- and the hotter extreme-He stars (EHe), or also the descendants of He-rich hot subdwarf stars (He-sdOs) \citep{2014A&A...566A.116R}. A possible scenario for these types of stars is the merger of two WDs \citep{1984ApJ...277..355W}. The merger of a C-O core WD with a He-core WD would produce an RCrB star or an EHe star \citep{2002MNRAS.333..121S,2011ApJ...737L..34L,2011MNRAS.414.3599J,2019MNRAS.488..438L}, meanwhile, the merger of two He-core WDs would produce a He-sdO \citep{2012MNRAS.419..452Z,2018MNRAS.476.5303S}. 
However, if the mass of the He-sdO is $M_{\star}\lesssim  0.5 M_{\odot}$, it can also be formed via a late hot-flasher scenario with a deep-mixing episode (as classified by \citealt{2004ApJ...602..342L}, see also \citealt{1993ApJ...407..649C,2001ApJ...562..368B,2003ApJ...582L..43C,2008A&A...491..253M}). 
This scenario is somehow similar to the VLTP, but instead of a late thermal pulse, what happens is a {\it late} onset of He-burning in the degenerate core of a low-mass star that has lost almost all -- but not all -- its H-rich envelope (having a $\sim 1$--5 $10^{-4}\, M_{\odot}$ envelope mass). 
In this scenario, the He flash occurs when the star has too low H-envelope mass to sustain a H-burning shell. Therefore, convection processes also carry H into the He-burning region where it is burned, leading to a H-deficient He-rich star. For this scenario to take place, the star needs to lose a significant amount of mass before the onset of He flash, in the red giant branch. The mass loss can occur due to the presence of a companion star, both via mass transfer due to stable Roche lobe overflow  or mass ejection in a common envelope phase \citep{1976IAUS...73...75P,2003MNRAS.341..669H}. Also, enhanced winds in the red giant branch, due, e.g., to rotation or enhanced initial He-compositions, could lead to a significant envelope-mass loss \citep{1997ApJ...474L..23S,2012ApJ...748...62V,2015Natur.523..318T,2017A&A...597A..67A}. Hereinafter, we will call this scenario the very-late hot-flasher (VLHF) scenario.
The outcome of the VLHF scenario would be a star with a mass necessarily close to the mass required to the onset of He burning in a degenerate He core, i.e., of about 0.45--0.49 $M_{\odot}$ -- depending on metallicity and He-abundance. 
 In summary, a low-mass ($\lesssim 0.5\,M_{\odot}$) H-deficient He-rich WD can come either from a VLTP of a low-mass star, a VLHF scenario or a merger of two low-mass He-core WDs. Now, the question arises as to whether there are H-deficient He-rich WDs with such masses.

Historically, DB WDs were found to have a mass distribution with a mean mass similar to the one of DA WDs ($\sim 0.6 M_{\odot}$), but without a significant spread to lower masses \citep{1979ApJ...228..240S,1984ApJ...281..276O,1996ASPC...96..295B,2001ApJS..133..413B,2007A&A...470.1079V,2011ApJ...737...28B}. 
However, some DB WDs in the range of 0.4--0.5 $M_{\odot}$ were found by different authors (e.g. \citealt{2015A&A...583A..86K} from a pre-$Gaia$ era). With the recent measurements of trigonometric parallaxes by $Gaia$, new mass distributions for WDs were derived in magnitude and volume-limited samples, for both H-rich and H-deficient WDs, using both photometric and spectroscopic techniques \citep{2019MNRAS.482..649O,2019ApJ...871..169G,2019ApJ...882..106G,2019MNRAS.482.5222T,2019MNRAS.482.4570G,2019ApJ...876...67B,2019MNRAS.486.2169K}. The most recent work for the case of He-rich WDs was carried out by \cite{2019ApJ...882..106G}. Regardless of the technique used, these authors found DB stars that appear to have masses below 0.5 $M_{\odot}$, even below 0.4 $M_{\odot}$. They argue that these stars are most likely double degenerate binaries (DB+DB) that are not resolved, and therefore, appear to have a large radius and hence a small mass. If this is not the case for all of them, however, they should have been formed through one of the scenarios mentioned above. Also, \cite{2014A&A...572A.117R} derived masses for a sample of DO WDs and found some of them with masses below 0.5 $M_{\odot}$. These authors argue that about $13$\% of the DO WDs may be the descendants of extreme horizontal branch (EHB) stars (i.e. He-sdO/B stars). 
All in all, the evolutionary history of DO and DB WDs is not completely clear. In particular, for DB/DO WDs with masses $\lesssim 0.5\, M_{\odot}$, a VLHF scenario for their formation is also a possibility. 

In this work, we aim to explore the differences in the evolutionary and pulsational properties of H-deficient low-mass WDs resulting from the VLTP and the VLHF scenarios, leaving the merger scenario for a future work. In order to do this, we take advantage of the fact that DB WDs are found to pulsate in the temperature range of $22\,000 {\rm K} < T_{\rm eff} < 30\,000\, {\rm K}$ \citep{2019A&ARv..27....7C}. They are called DBV or V777 Her variable stars. Asteroseismology is a powerful tool to explore the internal chemical stratification of stars \citep{2008PASP..120.1043F,2008ARA&A..46..157W,2010A&ARv..18..471A,2019A&ARv..27....7C}. The different physical processes taking place in the interior of stars that experience a VLTP or a VLHF would lead to different chemical profiles in the interior of the resultant low-mass WDs. These differences could have a distinct impact on the pulsational properties of the WDs. For instance, \cite{2017A&A...599A..21D} found differences in the period spectrum between DA WDs whose progenitors experienced thermal pulses and those DA WDs whose progenitors avoided the thermally pulsing phase. Motivated by this, we compare the pulsational properties of DB WDs models that come from these two scenarios (VLTP and the VLHF), with the aim of assessing the potential of asteroseismology to distinguish between those
scenarios. Finally, there exist in the literature detailed WD models evolved from PG1159 stars within the VLTP scenario \citep{2005A&A...435..631A,2006A&A...449..313M}; however, detailed models of WDs coming from He-sdO stars are lacking. We present here for the first time detailed WD models that come from He-sdO star models, within the VLHF scenario.
 
The paper is organized as follows. In Sect. 2 we introduce the evolutionary sequences for both the VLTP and the VLHF scenarios. In Sect. 3 we present the pulsational properties of WD models resulting from both scenarios and discuss their differences. In Sect. 4 we do a brief summary and present our conclusions.

\section{Evolutionary sequences}
\label{sec:evol}
\begin{figure*}
\centering
    \includegraphics[width=1\textwidth]{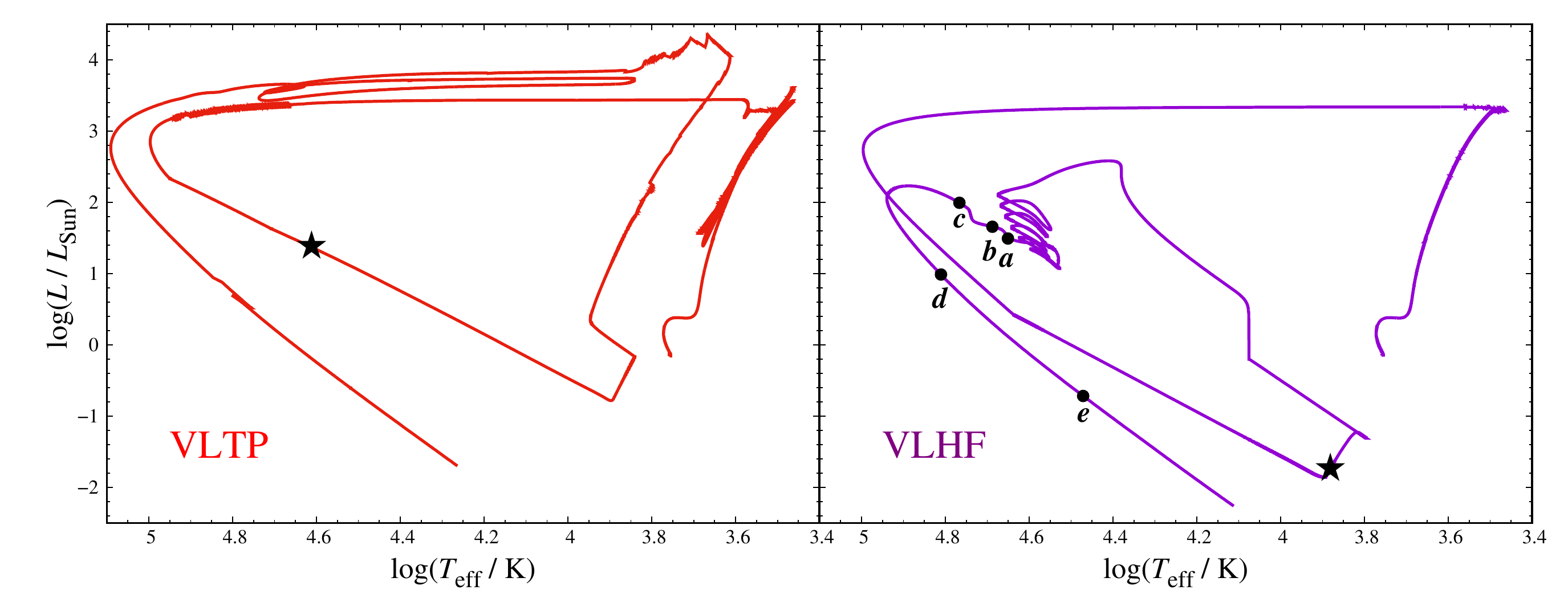}
  \caption{Hertzsprung-Russell diagram for the VLTP scenario (left) and the VLHF scenario (right) computed from the main sequence to the WD stage. In both panels, the star symbol marks the model where the maximum CNO-luminosity of the H flash occurs. In the right panel, we show the location of the models of Fig. \ref{fig:evol} ($a$, $b$, $c$, and $d$), and the model of the right panel of Fig. \ref{fig:profiles} ($e$).} 
  \label{fig:hr}
\end{figure*}

\begin{figure*}
\centering
    \includegraphics[width=1\textwidth]{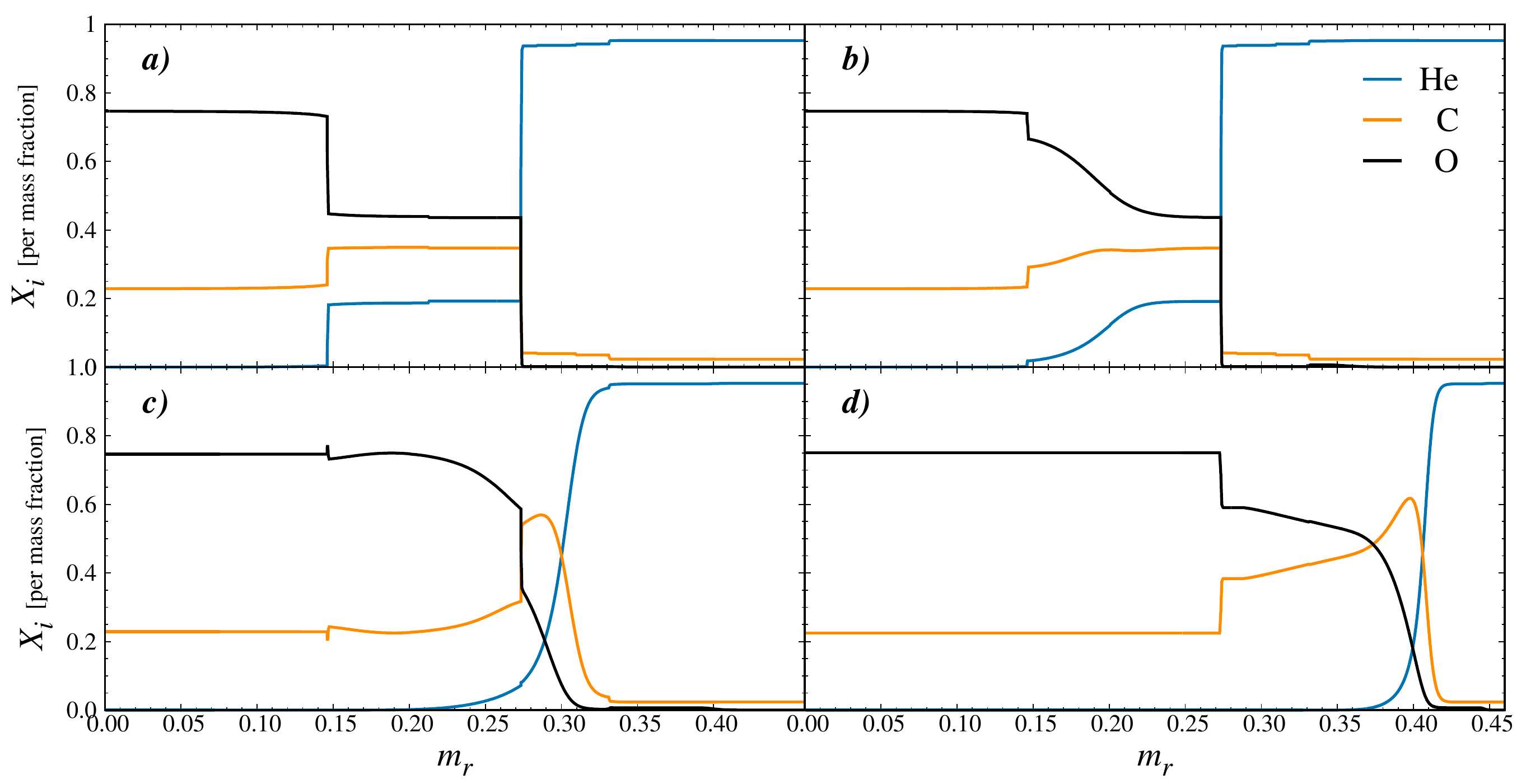}
  \caption{Oxygen (O), carbon (C) and helium (He) profiles of the models highlighted in Fig. \ref{fig:hr} with letters $a$, $b$, $c$ and $d$.} 
  \label{fig:evol}
\end{figure*}
For all the evolutionary calculations presented in this work we used the \texttt{LPCODE} stellar evolution code \citep{2005A&A...435..631A}. The most recent updates to the code can be found in \cite{2016A&A...588A..25M}. In Fig. \ref{fig:hr} we plot the Hertzsprung-Russell diagrams for both the VLTP and the VLHF sequences. In both sequences we mark the locus of the maximum CNO-luminosity of the H flash with a star symbol. 
The stellar masses predicted by the hot-flasher scenario depend on metallicity and range approximately from 0.455 $M_{\odot}$ for $Z = 0.02$ to 0.485 $M_{\odot}$ for $Z = 0.001$ \citep{2008A&A...491..253M,2018A&A...614A.136B}. We selected for this work a sequence of 0.46 $M_{\odot}$  with initial metallicity and He abundance of $Z= 0.02$ and $Y= 0.285$. This sequence was extracted from those calculated in the work of \cite{2018A&A...614A.136B}, and experiences a deep-mixing scenario \citep{2004ApJ...602..342L} where almost all the H is burned. We evolved this sequence to the WD regime. The VLTP sequence was computed from the ZAMS with the same initial metallicity and He abundance than the VLHF sequence, and an initial mass of $1\,M_{\odot}$. In the AGB phase we artificially removed mass of the star until the stellar mass was reduced to  $0.5\, M_{\odot}$, and continued the remnant evolution along two thermal pulses, forcing the last one to be a VLTP. The VLTP was followed by the corresponding H burning. Before the WD stage was reached, we relaxed the mass of the star to $0.46\, M_{\odot}$ in order to get rid of possible differences in the periods arising from differences in the mass of the models coming from the VLTP and the VLHF scenarios. For both sequences we used a MLT parameter of $\alpha_{\rm MLT}=1.822$, that corresponds to calibration of the solar model for the \texttt{LPCODE} -- see \citealt{2016A&A...588A..25M} -- , and overshooting in the central He-burning phase to an extent of $\sim 0.2$ the pressure scale height. The mass of the remaining H after the H flash was around $M_{\rm H}= 10^{-7}\,M_{\odot}$ in both cases.  For the VLHF this value drops below $10^{-10}\,M_{\odot}$ after the He subflashes. The exact value of the total amount of H that is burned in the H flash in both the VLTP and the VLHF scenarios depends on the details of convection and convective-border mixing processes. As the determination of the amount of convective-border mixing needed for obtaining a DB WD is beyond the scope of this work, we artificially removed the remaining H after the H flash in each case. 

\begin{figure*}
\centering
    \includegraphics[width=1\textwidth]{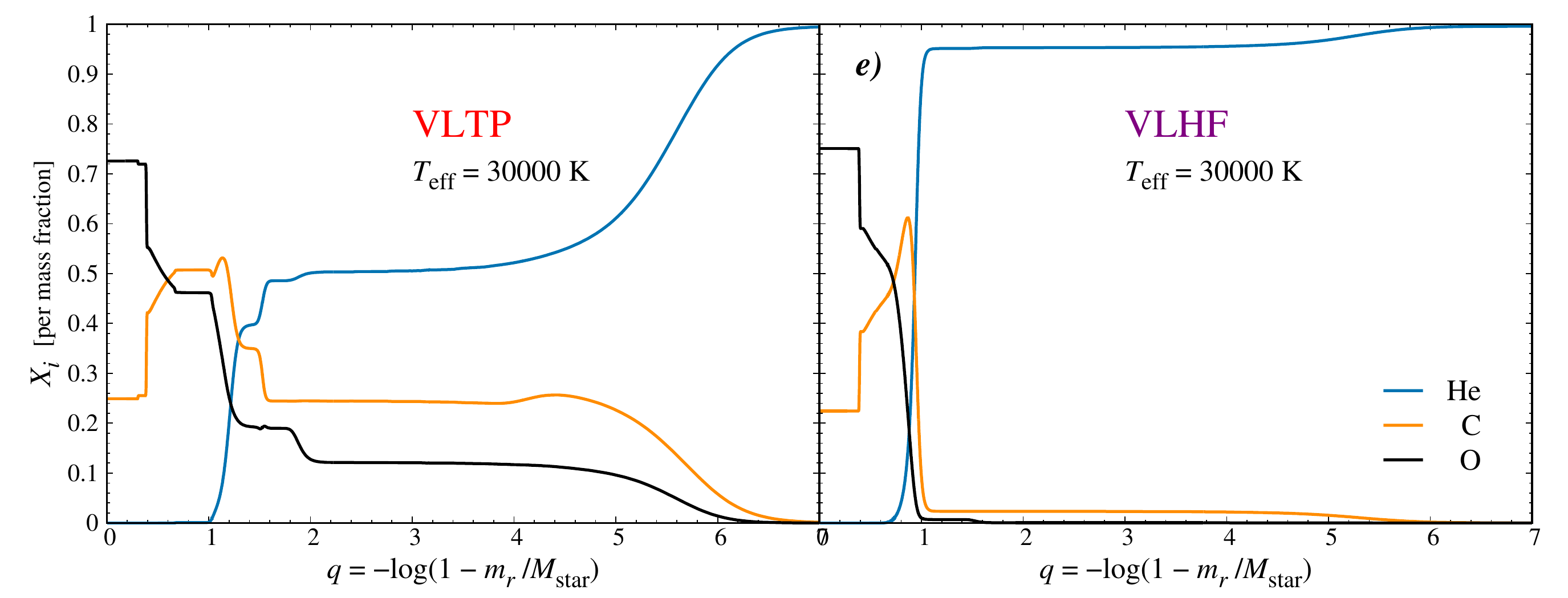}
  \caption{Oxygen (O), carbon (C) and helium (He) profiles of WD models coming from a VLTP (left) and a VLHF (right) at $T_{\rm eff}=30\,000\,$K.}
  \label{fig:profiles}
\end{figure*}

The evolution of PG1159 stars to the WD stage within the VLTP scenario is well documented \citep{2005A&A...435..631A,2006A&A...449..313M}. This is not the case for the evolution of He-sdOB stars to the WD stage within the VLHF scenario. We therefore show in Fig. \ref{fig:evol} the evolution of the chemical profile from the moment the star cease the central He-burning phase (panel $a$), to the WD stage (panel $d$). We show in Fig. \ref{fig:hr} the location of these models in the Hertzsprung-Russell diagram. In panel $a$ we can see the O-rich core up to $m_r\simeq0.15$. At this point, the central He-burning has ceased, but there is still He-burning in the layer between $m_r=0.15$ and $0.25$. In panels $b$ and $c$ we see how the He remaining in this layer is burned into C and O. In panel $d$ He-burning has finished, and the C/O core has settled to its final shape, as diffusion processes do not change significantly the chemical structure at the high temperatures of the core. This is already a WD model of $T_{\rm eff}=65800\,$K. Later, the model cools down to the DB WDs instability strip. In this cooling process, diffusion produces a pure He envelope as can be seen in the right panel of Fig. \ref{fig:profiles}, where we plot the WD model at $T_{\rm eff}=30\,000\,$K. The locus of this model in the Hertzsprung-Russell diagram is marked with an $e$ in Fig. \ref{fig:hr}.

We now compare the WD models coming from the VLTP and VLHF scenarios. In Fig. \ref{fig:profiles} we plot the chemical profiles of the WDs models coming from both the VLTP and the VLHF for $T_{\rm eff}=30\,000\,$K, that corresponds to the blue edge of the instability strip of the DBVs. In both evolutionary scenarios, the star evolves to a H-deficient WD, but the differences in the evolutionary history of both remnants are translated to different features in the chemical profiles. In the VLTP case, since thermal pulses have taken place, there are several changes in the slope of C and O profiles in the core. In the VLHF case, as no thermal pulse has taken place, the C/O profile has a simpler structure. Also, due to thermal pulses, in the VLTP model remains an intershell where a significant amount of He, C and O coexists, meanwhile in the VLHF model this feature is not present. Only a small amount of C up to $q = 5$ and O up to $q = 1.5$ can be seen as a consequence of the core He flash. In addition,  in the VLTP model,
where He-layer burning has been active more time than in the VLHF, the C/O core is more massive, being the C/He transition at $q \simeq 1.3$, in contrast with the value of $q=1$ for the C/He transition in the VLHF model.

\section{Pulsational properties}
\label{sec:results}

\begin{figure*}
\centering
    \includegraphics[width=1\textwidth]{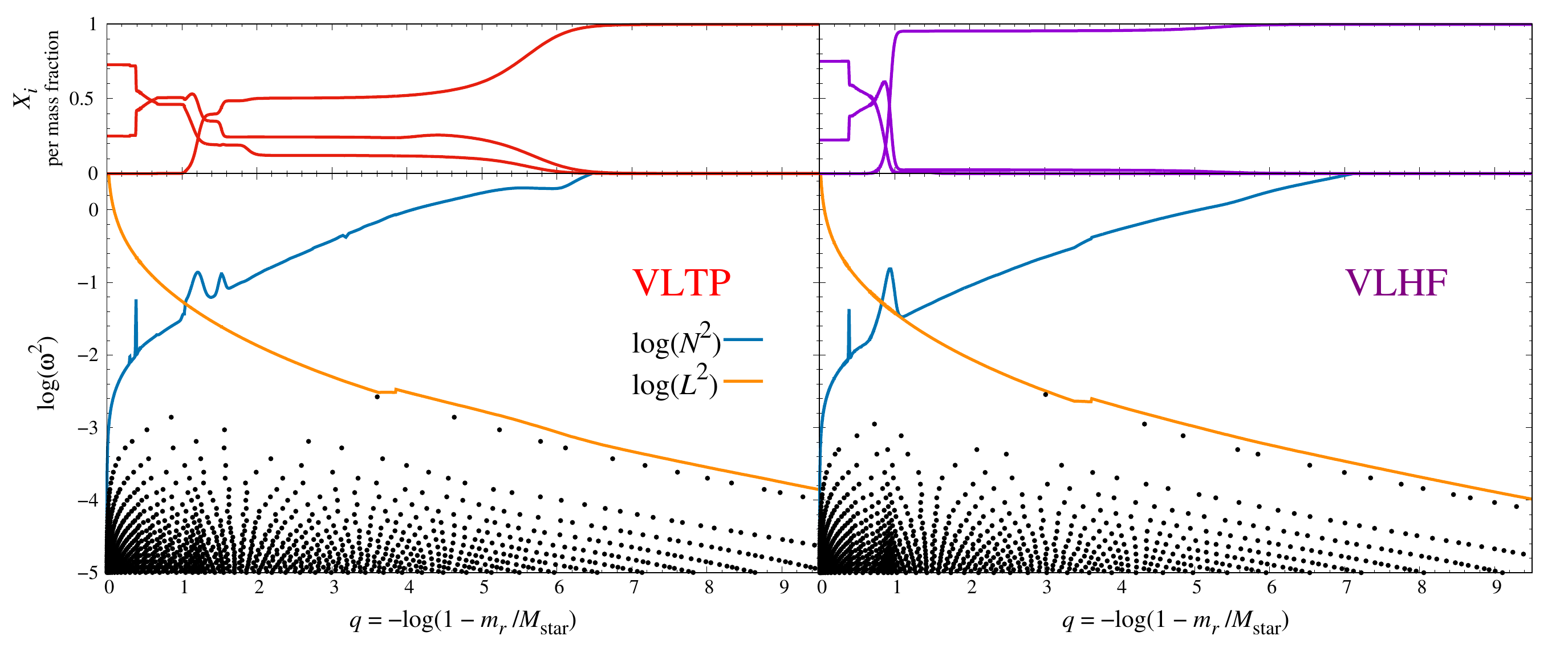}
  \caption{Lamb and Brunt-Väisälä frequencies (lower panels) and the chemical profile (upper panels) for the same models shown in Fig. \ref{fig:profiles}. In the lower panels, the black dots indicate the location of the nodes (zero displacement) of the radial eigenfunctions of $g$ modes.} 
  \label{fig:bv}
\end{figure*}

We calculated non-radial adiabatic gravity ($g$) modes for all the WDs models with effective temperatures in the range of the instability strip of DBVs ($22\,000\,\lesssim T_{\rm eff}\lesssim 32\,000\,$ K), for both the VLTP and VLHF scenarios. The periods observed in DB WDs range from 120 s to 1100 s approximately \citep{2019A&ARv..27....7C}. We calculated periods from 100 s to 2500 s, for harmonic degree $\ell = 1$ and $2$, thus covering the range of observed periods. All the calculations were made with the stellar-pulsation code \texttt{LP-PUL} \citep{2006A&A...458..259C}. We focus on pulsational results with $\ell= 1$ only because they qualitatively do not differ from the results for $\ell= 2$. In Fig. \ref{fig:bv} we show the logarithm of the squared Brunt-Väisälä and Lamb frequencies, plotted together with the chemical profiles for the same models of Fig. \ref{fig:profiles}. The Brunt-Väisälä frequency ($N$) is strongly dependent on the chemical structure. Any chemical interface imprints a bump in $N$. For this reason, for a WD that comes from a VLTP, $N$ has a more complicated structure than for a WD coming from a VLHF. In particular, we can see a bump at $q \sim 1.7$ in the VLTP model that is not present in the VLHF model because this last one lacks the intershell where a significant amount of O, C and He coexists. Also, the bump corresponding to the C-He transition at $q=1$ for the VLHF is more pronounced than the bump in the VLTP model. This is because the C-He transition in the VLHF profile is very well defined and steeper than in the VLTP case, making the  mode-trapping cavity of the core more noticeable in the VLHF profile. This cavity is also smaller than in the VLTP case, where the transition is at $q\simeq 1.3$. Therefore, we see differences in the Brunt-Väisälä frequency that arise from different features in the chemical profiles of the models. 
DBV WDs exhibit periods associated with $g$ modes. The properties of the $g$-mode spectrum are strongly dependent on the Brunt-Väisälä frequency. Therefore, we expect that the period spectrum of the DB WDs is affected by the differences in the Brunt-Väisälä frequencies of both models. We discuss in the following sections the impact of these differences on the period spectrum of the models, the period separations and the period drifts.

\subsection{Periods}
\label{sec:periods}

\begin{figure*}
\centering
    \includegraphics[width=1\textwidth]{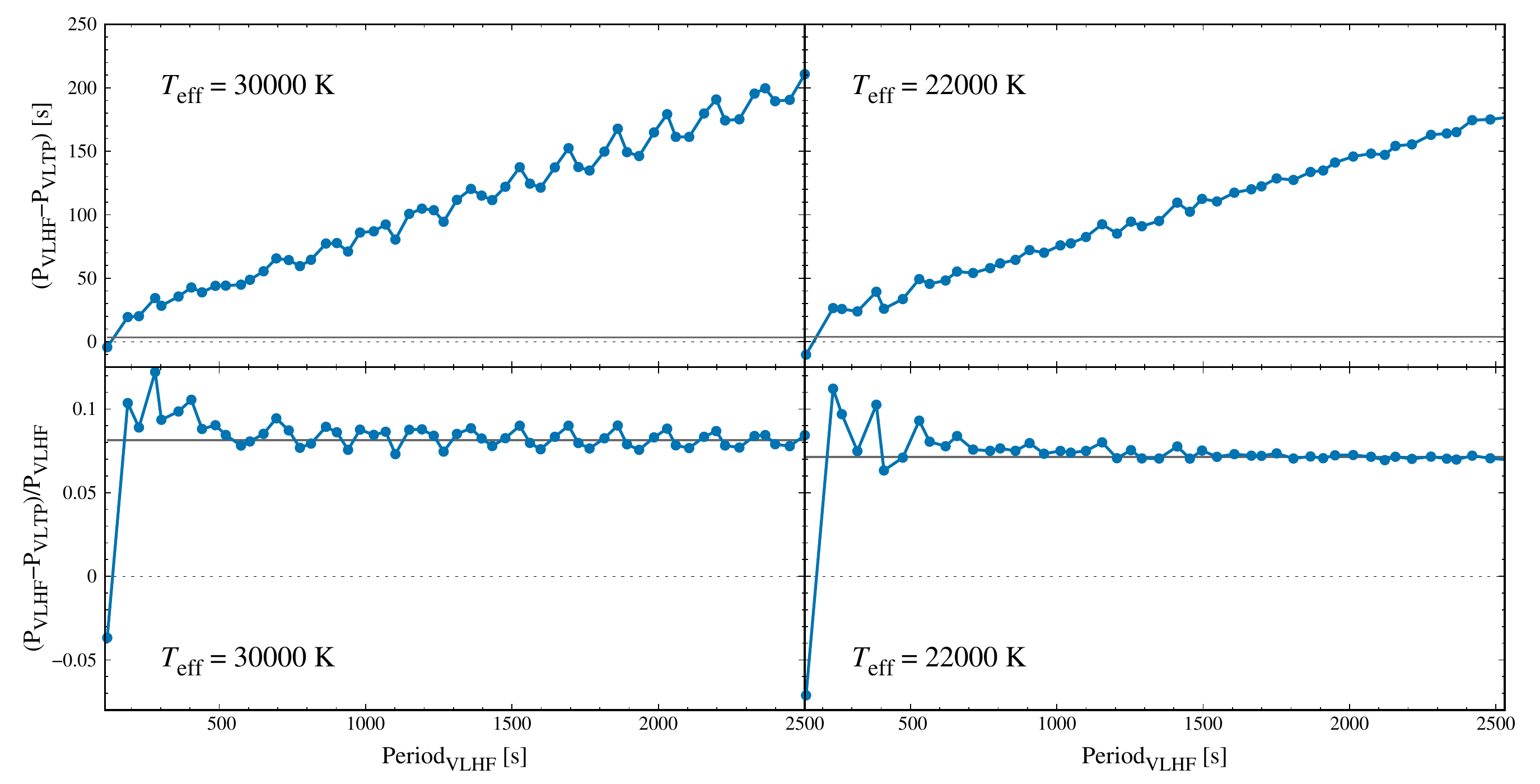}
  \caption{Difference between periods (blue lines) and asymptotic period spacings (horizontal full gray lines) of the VLHF and the VLTP models with the same radial order $k$ vs. the periods of the VLHF models for $T_{\rm eff}=30\,000\,$K and 22\,000$\,$K (upper panels), and the same differences but relative to the periods of the VLHF models (lower panels).} 
  \label{fig:periods}
\end{figure*}

In the upper panels of Fig. \ref{fig:periods} we show the differences between the periods of the VLHF and VLTP models for $T_{\rm eff}$ values of $30\,000\,$K and $22\,000\,$K. In the lower panels, we show those differences relative to the periods of the VLHF models for the same temperatures. All the differences are between periods with the same radial order $k$. In the case of $g$ modes, lower radial orders correspond to shorter periods. \cite{2017A&A...599A..21D} found that the differences in periods of a pulsating H-rich WD (DAV WD) model that has experienced 3 thermal pulses respect to one that has not experienced thermal pulses are less than $15\,$s for $0.548\,M_{\odot}$ in the period range of DA WD stars, at $T_{\rm eff}=12000\,$K. In contrast with this result, for our H-deficient WDs, we find that the differences of periods between no thermal pulses (VLHF) and two thermal pulses (VLTP) can be as high as $100\,$s for $k=23$ -- corresponding to a period of $1080\,$s for the VLHF -- and grows with higher values of $k$. This is probably due to the differences in the asymptotic period spacing of the models -- see Sect. \ref{sec:period_spacings}. Also, we find that the periods for the VLHF models are systematically higher than the periods of the VLTP models, except for $k= 1$. All the differences are due to both the existence of the intershell region and the shift outward of the C-He transition region in the case of the VLTP models, in comparison with the VLHF case. The higher differences compared to the ones found for H-rich pulsating WDs by \cite{2017A&A...599A..21D} may be related with the fact that we are comparing a model that went trough the AGB and the thermally-pulsing phase with one that avoided the AGB phase. In \cite{2017A&A...599A..21D}, all the models that they compare went through the AGB phase, even if they avoided the thermally-pulsing phase. Therefore, the chemical profiles that they compare show less-pronounced differences than the ones that we compare in this work. Also, the different results may  be related to the difference in mass and temperature of our models respect to the ones of \cite{2017A&A...599A..21D}. More massive WDs are more dense stars (because of the mass-radius relation for WDs), and therefore, the Brunt-Väisälä frequency is higher. This means that the period spectrum of more massive models moves to shorter values, compared to less massive models. As a consequence, absolute differences in periods when comparing higher-mass models are expected to be lower than in lower-mass models. In addition, diffusion processes are still very active at the temperatures of pulsating DB stars, and the differences in the chemical profile are more pronounced than at lower temperatures ($T_{\rm eff}\sim 12\,000\,$ K) where DAV stars pulsate.

Though the higher differences are found for longer periods -- and higher values of $k$ -- , the higher relative differences are found for periods with radial order up to $k=10$, being the highest differences for $k= 4$ -- about $12\%$, see Fig. \ref{fig:periods}, lower panel. The horizontal line in the plots is the relative difference between the asymptotic period spacings. The relative differences in periods vary around this value, specially for high values of $k$. This is showing that the differences at long periods ($P > 600\,$s) are mainly due to the differences in the asymptotic period spacings, and the differences at short periods ($P < 600\,$s) are due also to the differences in the chemical profiles.

The important differences found in the periods of VLHF and VLTP models of the same mass and temperatures suggest that it might be possible to infer valuable information about  the evolutionary history of low-mass DB WDs by means of asteroseismic period-to-period fits of DBV stars.

\subsection{Period spacings}
\label{sec:period_spacings}

\begin{figure*}
\centering
    \includegraphics[width=1\textwidth]{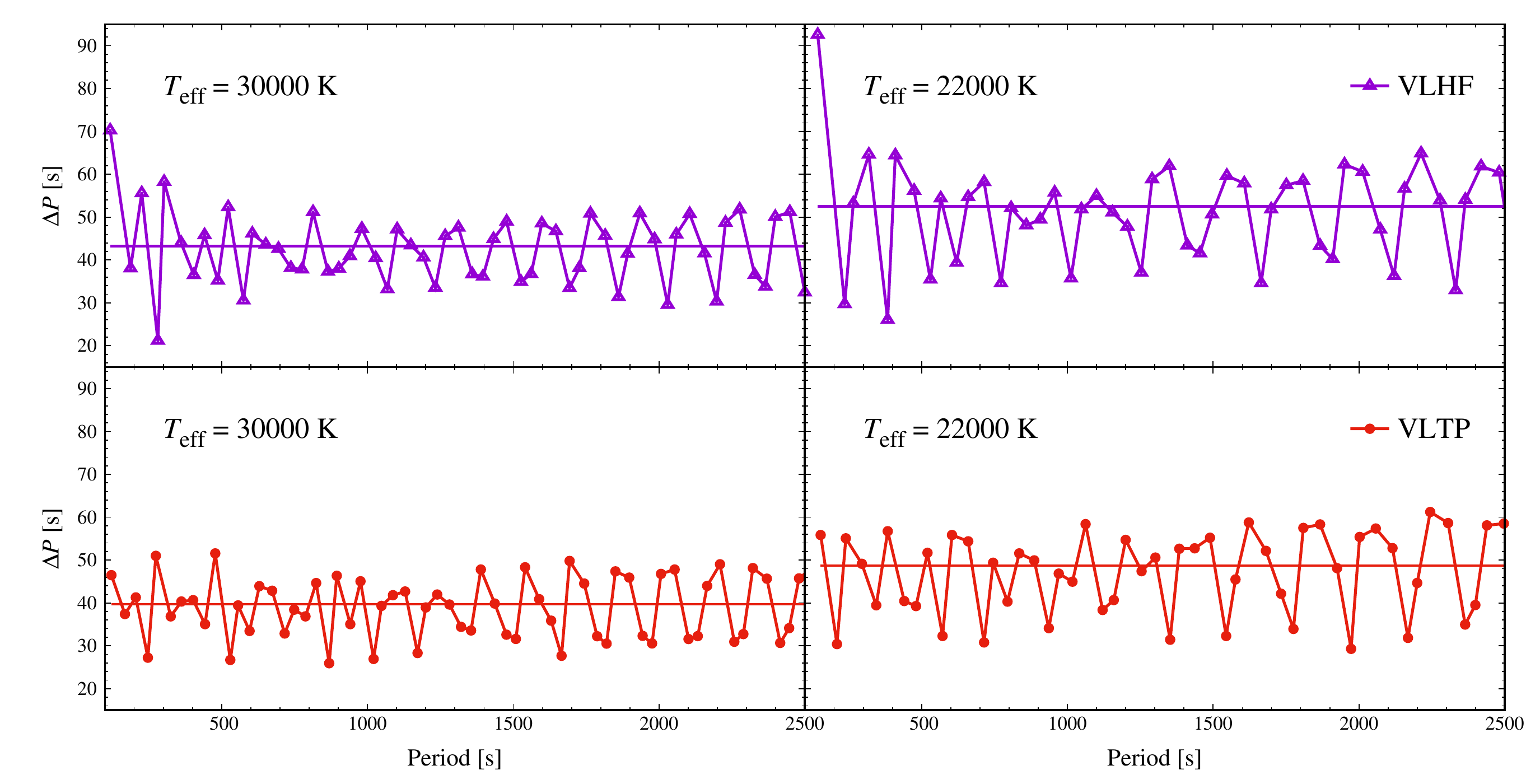}
  \caption{Period spacings ($\Delta P$) vs. periods for the VLTP case (red lines, lower panels) and the VLHF case (violet lines, upper panels), for $T_{\rm eff}= 30\,000\,$K (right panels) and $T_{\rm eff}= 22\,000\,$K (left panels). Horizontal lines show the corresponding asymptotic period spacings.} 
  \label{fig:deltap}
\end{figure*}

In Fig. \ref{fig:deltap} we show the period spacings $\Delta P$, which are the differences between consecutive periods of a same model. We compare the periods spacings for both evolutionary scenarios at $T_{\rm eff}= 30\,000\,$K and $22\,000\,$K. We plot, for comparison, the asymptotic period spacings ($\Delta P$ when $k\,\rightarrow \infty$). In all cases, the period-spacing distribution show mode-trapping substructures, that change as diffusion acts. However, these trapping substructures are somewhat different for the two scenarios, in particular the trapping amplitudes. At both values of $T_{\rm eff}$, the trapping amplitude of the VLHF model is higher for periods between $\simeq$ 100--400 s. In fact, we already discussed in Sect. \ref{sec:periods} that this range of periods is the most affected by the differences in the chemical profiles. Therefore, measuring period spacings between periods in the range of 100-400 s also can help us to determine the evolutionary history of low-mass DB WDs.

The Brunt-Väisälä frequency of the VLTP model is somewhat higher than in the VLHF model, above $q \sim 0.5$. This is due to the existence of the intershell region and the location of the C-He transition region in the VLTP model. The differences on $N$ in both models make the asymptotic period spacings to slightly differ, on about $2$--$3\,$s for all temperatures. Therefore, even if we were able to measure sufficient periods in a pulsating low-mass DB WD in order to determine a mean period spacing, that would not be likely useful for distinguishing between the two scenarios.

\subsection{Period drifts}

\begin{figure*}
\centering
    \includegraphics[width=1\textwidth]{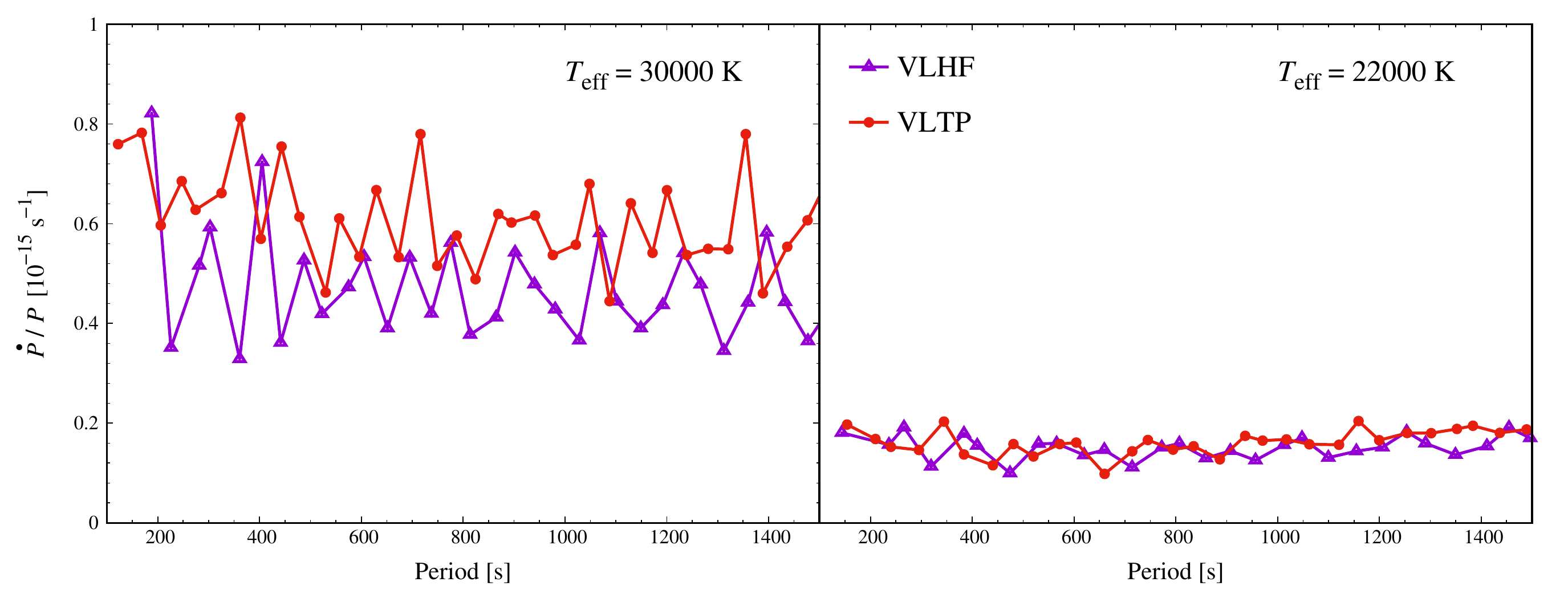}
  \caption{Period drifts ($\dot P$) relative to periods ($P$) vs. periods for the VLTP case (red lines) and the VLHF case (violet lines), for $T_{\rm eff}=30\,000\,$K (right panel) and $T_{\rm eff}=22\,000\,$K (left panel).} 
  \label{fig:dotper}
\end{figure*}

Another observable quantity from pulsations in WDs is the rate of change of periods (period drifts, $\dot P$). Due to the difficulty in finding stable periods in DBVs, a reliable measurement of the period drift for a DB WD is lacking. However, \cite{2011MNRAS.415.1220R} have derived an estimate of the period drift for the DBV star PG 1351+489. We calculated the period drifts of the models to see if future determinations of $\dot P$ in DBVs could help us to determine their evolutionary history. In Fig. \ref{fig:dotper} we plot the period drifts relative to periods of the same models shown in Figs. \ref{fig:periods} and \ref{fig:deltap}. We can see that for $T_{\rm eff}= 30\,000\,$ K, the mean value of the period drifts for the VLHF model is larger than for the VLTP model, but this is not true for all the periods. Therefore, determining one period drift is not likely to help us to distinguish between the models. In the case of $T_{\rm eff}=22\,000\,$ K there are almost no differences in the values of $\dot P$. The period drift of both DAVs and DBVs are related to their cooling rates \citep{2010A&ARv..18..471A}. At $T_{\rm eff}= 30\,000\,$K the VLHF model evolves faster than the VLTP model, but this is not the case shortly after, therefore, we do not expect significant differences in $\dot P$ for temperatures below  $T_{\rm eff}= 30\,000\,$ K. As the period drifts for the models of VLHF and VLTP do not exhibit significant differences, measuring a period drift of low-mass DB WDs will not help us in distinguish between VLTP and VLHF models -- unless we were able to measure the period drift for several periods of a star with $T_{\rm eff}\sim 30\,000\,$K.

\section{Discussion and conclusions}
\label{sec:disc}

In this work, we compared the pulsations properties of low-mass DB WD models that came from a very-late thermal pulse after the AGB phase with models that experienced a late helium flash onset in an almost stripped core of a red giant star. We find that these two evolutionary channels for the formation of low-mass ($\sim 0.46\,M_{\odot}$) H-deficient WDs lead to very different chemical profiles at the WD stage. As a consequence, the Brunt-Väisälä frequencies of the models at the instability strip of DBVs exhibit different features that translate into different properties of the $g$-mode pulsation spectrum. In particular, the periods in the range of 100--400 s are more sensitive to the distinct features of the chemical profiles, showing different mode-trapping substructures. This implies that both period-to-period fits of DB WDs and the measurement of the period spacings in the mentioned range of periods can help to determine the evolutionary history of low-mass DB WDs. In contrast, mean period spacings and period drifts measurements are not likely to help to distinguish between the two evolutionary scenarios. These last two quantities are more complicated to determine for DBVs, because of the low number of periods usually observed in WDs, and the difficulty in finding stable periods in pulsating DB WDs in particular.
Therefore, we conclude that both, a comprehensive analysis of the observed period spacings in pulsating low-mass DB WDs, especially in the range of periods below 500$\,$s, and detailed asteroseismic period-to-period fits of these stars could help to shed light about their evolutionary history.

\begin{acknowledgements}
We cordially thank the anonymous referee for a constructive report that improved the presentation of this work. We thank Marcelo Miller Bertolami for useful discussions. 
Part of this work was supported by AGENCIA through the Programa de Modernización Tecnológica BID 1728/OC-AR, by the PIP 112-200801-00940 grant from CONICET and by the G149 grant from the National University of La Plata. This research has made use of NASA's Astrophysics Data System.
\end{acknowledgements}

   \bibliographystyle{aa} 
   \bibliography{bibliografia} 

\begin{thebibliography}{50}
\expandafter\ifx\csname natexlab\endcsname\relax\def\natexlab#1{#1}\fi

\bibitem[{{Althaus} {et~al.}(2010){Althaus}, {C{\'o}rsico}, {Isern}, \&
  {Garc{\'{\i}}a-Berro}}]{2010A&ARv..18..471A}
{Althaus}, L.~G., {C{\'o}rsico}, A.~H., {Isern}, J., \& {Garc{\'{\i}}a-Berro},
  E. 2010, \aapr, 18, 471

\bibitem[{{Althaus} {et~al.}(2017){Althaus}, {De Ger{\'o}nimo}, {C{\'o}rsico},
  {Torres}, \& {Garc{\'{\i}}a-Berro}}]{2017A&A...597A..67A}
{Althaus}, L.~G., {De Ger{\'o}nimo}, F., {C{\'o}rsico}, A., {Torres}, S., \&
  {Garc{\'{\i}}a-Berro}, E. 2017, \aap, 597, A67

\bibitem[{{Althaus} {et~al.}(2005){Althaus}, {Serenelli}, {Panei},
  {C{\'o}rsico}, {Garc{\'\i}a-Berro}, \& {Sc{\'o}ccola}}]{2005A&A...435..631A}
{Althaus}, L.~G., {Serenelli}, A.~M., {Panei}, J.~A., {et~al.} 2005, \aap, 435,
  631

\bibitem[{{Battich} {et~al.}(2018){Battich}, {Miller Bertolami}, {C{\'o}rsico},
  \& {Althaus}}]{2018A&A...614A.136B}
{Battich}, T., {Miller Bertolami}, M.~M., {C{\'o}rsico}, A.~H., \& {Althaus},
  L.~G. 2018, \aap, 614, A136

\bibitem[{{Beauchamp} {et~al.}(1996){Beauchamp}, {Wesemael}, {Bergeron},
  {Liebert}, \& {Saffer}}]{1996ASPC...96..295B}
{Beauchamp}, A., {Wesemael}, F., {Bergeron}, P., {Liebert}, J., \& {Saffer},
  R.~A. 1996, in Astronomical Society of the Pacific Conference Series,
  Vol.~96, Hydrogen Deficient Stars, ed. C.~S. {Jeffery} \& U.~{Heber}, 295

\bibitem[{{Bergeron} {et~al.}(2019){Bergeron}, {Dufour}, {Fontaine}, {Coutu},
  {Blouin}, {Genest-Beaulieu}, {B{\'e}dard}, \& {Rolland
  }}]{2019ApJ...876...67B}
{Bergeron}, P., {Dufour}, P., {Fontaine}, G., {et~al.} 2019, \apj, 876, 67

\bibitem[{{Bergeron} {et~al.}(2001){Bergeron}, {Leggett}, \&
  {Ruiz}}]{2001ApJS..133..413B}
{Bergeron}, P., {Leggett}, S.~K., \& {Ruiz}, M.~T. 2001, \apjs, 133, 413

\bibitem[{{Bergeron} {et~al.}(2011){Bergeron}, {Wesemael}, {Dufour},
  {Beauchamp}, {Hunter}, {Saffer}, {Gianninas}, {Ruiz}, {Limoges}, {Dufour},
  {Fontaine}, \& {Liebert}}]{2011ApJ...737...28B}
{Bergeron}, P., {Wesemael}, F., {Dufour}, P., {et~al.} 2011, \apj, 737, 28

\bibitem[{{Brown} {et~al.}(2001){Brown}, {Sweigart}, {Lanz}, {Landsman}, \&
  {Hubeny}}]{2001ApJ...562..368B}
{Brown}, T.~M., {Sweigart}, A.~V., {Lanz}, T., {Landsman}, W.~B., \& {Hubeny},
  I. 2001, \apj, 562, 368

\bibitem[{{Cassisi} {et~al.}(2003){Cassisi}, {Schlattl}, {Salaris}, \&
  {Weiss}}]{2003ApJ...582L..43C}
{Cassisi}, S., {Schlattl}, H., {Salaris}, M., \& {Weiss}, A. 2003, \apjl, 582,
  L43

\bibitem[{{Castellani} \& {Castellani}(1993)}]{1993ApJ...407..649C}
{Castellani}, M. \& {Castellani}, V. 1993, \apj, 407, 649

\bibitem[{{C{\'o}rsico} {et~al.}(2006){C{\'o}rsico}, {Althaus}, \& {Miller
  Bertolami}}]{2006A&A...458..259C}
{C{\'o}rsico}, A.~H., {Althaus}, L.~G., \& {Miller Bertolami}, M.~M. 2006,
  \aap, 458, 259

\bibitem[{{C{\'o}rsico} {et~al.}(2019){C{\'o}rsico}, {Althaus}, {Miller
  Bertolami}, \& {Kepler}}]{2019A&ARv..27....7C}
{C{\'o}rsico}, A.~H., {Althaus}, L.~G., {Miller Bertolami}, M.~M., \& {Kepler},
  S.~O. 2019, \aapr, 27, 7

\bibitem[{{De Ger{\'o}nimo} {et~al.}(2017){De Ger{\'o}nimo}, {Althaus},
  {C{\'o}rsico}, {Romero}, \& {Kepler}}]{2017A&A...599A..21D}
{De Ger{\'o}nimo}, F.~C., {Althaus}, L.~G., {C{\'o}rsico}, A.~H., {Romero},
  A.~D., \& {Kepler}, S.~O. 2017, \aap, 599, A21

\bibitem[{{Fontaine} \& {Brassard}(2008)}]{2008PASP..120.1043F}
{Fontaine}, G. \& {Brassard}, P. 2008, \pasp, 120, 1043

\bibitem[{{Genest-Beaulieu} \&
  {Bergeron}(2019{\natexlab{a}})}]{2019ApJ...871..169G}
{Genest-Beaulieu}, C. \& {Bergeron}, P. 2019{\natexlab{a}}, \apj, 871, 169

\bibitem[{{Genest-Beaulieu} \&
  {Bergeron}(2019{\natexlab{b}})}]{2019ApJ...882..106G}
---. 2019{\natexlab{b}}, \apj, 882, 106

\bibitem[{{Gentile Fusillo} {et~al.}(2019){Gentile Fusillo}, {Tremblay},
  {G{\"a}nsicke}, {Manser}, {Cunningham}, {Cukanovaite}, {Hollands}, {Marsh},
  {Raddi}, {Jordan}, {Toonen}, {Geier}, {Barstow}, \&
  {Cummings}}]{2019MNRAS.482.4570G}
{Gentile Fusillo}, N.~P., {Tremblay}, P.-E., {G{\"a}nsicke}, B.~T., {et~al.}
  2019, \mnras, 482, 4570

\bibitem[{{Han} {et~al.}(2003){Han}, {Podsiadlowski}, {Maxted}, \&
  {Marsh}}]{2003MNRAS.341..669H}
{Han}, Z., {Podsiadlowski}, P., {Maxted}, P.~F.~L., \& {Marsh}, T.~R. 2003,
  \mnras, 341, 669

\bibitem[{{Iben}(1976)}]{1976ApJ...208..165I}
{Iben}, Jr., I. 1976, \apj, 208, 165

\bibitem[{{Iben} {et~al.}(1983){Iben}, {Kaler}, {Truran}, \&
  {Renzini}}]{1983ApJ...264..605I}
{Iben}, Jr., I., {Kaler}, J.~B., {Truran}, J.~W., \& {Renzini}, A. 1983, \apj,
  264, 605

\bibitem[{{Jeffery} {et~al.}(2011){Jeffery}, {Karakas}, \&
  {Saio}}]{2011MNRAS.414.3599J}
{Jeffery}, C.~S., {Karakas}, A.~I., \& {Saio}, H. 2011, \mnras, 414, 3599

\bibitem[{{Kepler} {et~al.}(2019){Kepler}, {Pelisoli}, {Koester}, {Reindl},
  {Geier}, {Romero}, {Ourique}, {Oliveira}, \& {Amaral}}]{2019MNRAS.486.2169K}
{Kepler}, S.~O., {Pelisoli}, I., {Koester}, D., {et~al.} 2019, \mnras, 486,
  2169

\bibitem[{{Koester} \& {Kepler}(2015)}]{2015A&A...583A..86K}
{Koester}, D. \& {Kepler}, S.~O. 2015, \aap, 583, A86

\bibitem[{{Lanz} {et~al.}(2004){Lanz}, {Brown}, {Sweigart}, {Hubeny}, \&
  {Landsman}}]{2004ApJ...602..342L}
{Lanz}, T., {Brown}, T.~M., {Sweigart}, A.~V., {Hubeny}, I., \& {Landsman},
  W.~B. 2004, \apj, 602, 342

\bibitem[{{Lauer} {et~al.}(2019){Lauer}, {Chatzopoulos}, {Clayton}, {Frank}, \&
  {Marcello}}]{2019MNRAS.488..438L}
{Lauer}, A., {Chatzopoulos}, E., {Clayton}, G.~C., {Frank}, J., \& {Marcello},
  D.~C. 2019, \mnras, 488, 438

\bibitem[{{Longland} {et~al.}(2011){Longland}, {Lor{\'e}n-Aguilar}, {Jos{\'e}},
  {Garc{\'{\i}}a-Berro}, {Althaus}, \& {Isern}}]{2011ApJ...737L..34L}
{Longland}, R., {Lor{\'e}n-Aguilar}, P., {Jos{\'e}}, J., {et~al.} 2011, \apjl,
  737, L34

\bibitem[{{Miller Bertolami}(2016)}]{2016A&A...588A..25M}
{Miller Bertolami}, M.~M. 2016, \aap, 588, A25

\bibitem[{{Miller Bertolami} {et~al.}(2006){Miller Bertolami}, {Althaus},
  {Serenelli}, \& {Panei}}]{2006A&A...449..313M}
{Miller Bertolami}, M.~M., {Althaus}, L.~G., {Serenelli}, A.~M., \& {Panei},
  J.~A. 2006, \aap, 449, 313

\bibitem[{{Miller Bertolami} {et~al.}(2008){Miller Bertolami}, {Althaus},
  {Unglaub}, \& {Weiss}}]{2008A&A...491..253M}
{Miller Bertolami}, M.~M., {Althaus}, L.~G., {Unglaub}, K., \& {Weiss}, A.
  2008, \aap, 491, 253

\bibitem[{{Oke} {et~al.}(1984){Oke}, {Weidemann}, \&
  {Koester}}]{1984ApJ...281..276O}
{Oke}, J.~B., {Weidemann}, V., \& {Koester}, D. 1984, \apj, 281, 276

\bibitem[{{Ourique} {et~al.}(2019){Ourique}, {Romero}, {Kepler}, {Koester}, \&
  {Amaral}}]{2019MNRAS.482..649O}
{Ourique}, G., {Romero}, A.~D., {Kepler}, S.~O., {Koester}, D., \& {Amaral},
  L.~A. 2019, \mnras, 482, 649

\bibitem[{{Paczynski}(1976)}]{1976IAUS...73...75P}
{Paczynski}, B. 1976, in IAU Symposium, Vol.~73, Structure and Evolution of
  Close Binary Systems, ed. P.~{Eggleton}, S.~{Mitton}, \& J.~{Whelan}, 75

\bibitem[{{Redaelli} {et~al.}(2011){Redaelli}, {Kepler}, {Costa}, {Winget},
  {Handler}, {Castanheira}, {Kanaan}, {Fraga}, {Henrique}, {Giovannini},
  {Provencal}, {Shipman}, {Dalessio}, {Thompson}, {Mullally}, {Brewer},
  {Childers}, {Oksala}, {Rosen}, {Wood}, {Reed}, {Walter}, {Strickland},
  {Chandler}, {Watson}, {Nather}, {Montgomery}, {Bischoff-Kim}, {Hansen},
  {Nitta}, {Kleinman}, {Claver}, {Brown}, {Sullivan}, {Kim}, {Chen}, {Yang},
  {Shih}, {Zhang}, {Jiang}, {Fu}, {Seetha}, {Ashoka}, {Marar}, {Baliyan},
  {Vats}, {Chernyshev}, {Ibbetson}, {Leibowitz}, {Hemar}, {Sergeev}, {Andreev},
  {Janulis}, {Mei{\v{s}}tas}, {Moskalik}, {Pajdosz}, {Baran}, {Winiarski},
  {Zola}, {Ogloza}, {Siwak}, {Bogn{\'a}r}, {Solheim}, {Sefako}, {Buckley},
  {O'Donoghue}, {Nagel}, {Silvotti}, {Bruni}, {Fremy}, {Vauclair}, {Chevreton},
  {Dolez}, {Pfeiffer}, {Barstow}, {Creevey}, {Kawaler}, \&
  {Clemens}}]{2011MNRAS.415.1220R}
{Redaelli}, M., {Kepler}, S.~O., {Costa}, J.~E.~S., {et~al.} 2011, \mnras, 415,
  1220

\bibitem[{{Reindl} {et~al.}(2014{\natexlab{a}}){Reindl}, {Rauch}, {Werner},
  {Kepler}, {G{\"a}nsicke}, \& {Gentile Fusillo}}]{2014A&A...572A.117R}
{Reindl}, N., {Rauch}, T., {Werner}, K., {et~al.} 2014{\natexlab{a}}, \aap,
  572, A117

\bibitem[{{Reindl} {et~al.}(2014{\natexlab{b}}){Reindl}, {Rauch}, {Werner},
  {Kruk}, \& {Todt}}]{2014A&A...566A.116R}
{Reindl}, N., {Rauch}, T., {Werner}, K., {Kruk}, J.~W., \& {Todt}, H.
  2014{\natexlab{b}}, \aap, 566, A116

\bibitem[{{Renzini}(1979)}]{1979ASSL...75..155R}
{Renzini}, A. 1979, in Astrophysics and Space Science Library, Vol.~75, Stars
  and star systems, ed. B.~E. {Westerlund}, 155--171

\bibitem[{{Renzini}(1981)}]{1981ASSL...89..319R}
{Renzini}, A. 1981, in Astrophysics and Space Science Library, Vol.~89, IAU
  Colloq. 59: Effects of Mass Loss on Stellar Evolution, ed. C.~{Chiosi} \&
  R.~{Stalio}, 319--336

\bibitem[{{Saio} \& {Jeffery}(2002)}]{2002MNRAS.333..121S}
{Saio}, H. \& {Jeffery}, C.~S. 2002, \mnras, 333, 121

\bibitem[{{Schoenberner}(1979)}]{1979A&A....79..108S}
{Schoenberner}, D. 1979, \aap, 79, 108

\bibitem[{{Schwab}(2018)}]{2018MNRAS.476.5303S}
{Schwab}, J. 2018, \mnras, 476, 5303

\bibitem[{{Shipman}(1979)}]{1979ApJ...228..240S}
{Shipman}, H.~L. 1979, \apj, 228, 240

\bibitem[{{Sweigart}(1997)}]{1997ApJ...474L..23S}
{Sweigart}, A.~V. 1997, \apjl, 474, L23

\bibitem[{{Tailo} {et~al.}(2015){Tailo}, {D'Antona}, {Vesperini}, {di
  Criscienzo}, {Ventura}, {Milone}, {Bellini}, {Dotter}, {Decressin},
  {D'Ercole}, {Caloi}, \& {Capuzzo-Dolcetta}}]{2015Natur.523..318T}
{Tailo}, M., {D'Antona}, F., {Vesperini}, E., {et~al.} 2015, \nat, 523, 318

\bibitem[{{Tremblay} {et~al.}(2019){Tremblay}, {Cukanovaite}, {Gentile
  Fusillo}, {Cunningham}, \& {Hollands}}]{2019MNRAS.482.5222T}
{Tremblay}, P.~E., {Cukanovaite}, E., {Gentile Fusillo}, N.~P., {Cunningham},
  T., \& {Hollands}, M.~A. 2019, \mnras, 482, 5222

\bibitem[{{Villanova} {et~al.}(2012){Villanova}, {Geisler}, {Piotto}, \&
  {Gratton}}]{2012ApJ...748...62V}
{Villanova}, S., {Geisler}, D., {Piotto}, G., \& {Gratton}, R.~G. 2012, \apj,
  748, 62

\bibitem[{{Voss} {et~al.}(2007){Voss}, {Koester}, {Napiwotzki}, {Christlieb},
  \& {Reimers}}]{2007A&A...470.1079V}
{Voss}, B., {Koester}, D., {Napiwotzki}, R., {Christlieb}, N., \& {Reimers}, D.
  2007, \aap, 470, 1079

\bibitem[{{Webbink}(1984)}]{1984ApJ...277..355W}
{Webbink}, R.~F. 1984, \apj, 277, 355

\bibitem[{{Winget} \& {Kepler}(2008)}]{2008ARA&A..46..157W}
{Winget}, D.~E. \& {Kepler}, S.~O. 2008, \araa, 46, 157

\bibitem[{{Zhang} \& {Jeffery}(2012)}]{2012MNRAS.419..452Z}
{Zhang}, X. \& {Jeffery}, C.~S. 2012, \mnras, 419, 452

\end{thebibliography}

\end{document}